\documentstyle[multicol,prl,aps,graphics]{revtex}
\begin{document}

\title{
Tuning Fermi-surface properties through quantum confinement in
metallic meta-lattices: New metals from old atoms}
\author{J. E. Han and Vincent H. Crespi}
\address{Department of Physics and the Center for Materials Physics, 
The Pennsylvania State University, 104 Davey Lab, University Park, PA,
16802-6300; vhc2@psu.edu}
\date{\today}
\maketitle

\begin{abstract}
We describe a new class of nanoscale structured metals wherein the effects
of quantum confinement are combined with dispersive metallic electronic
states to induce modifications to the fundamental low-energy microscopic
properties of a three-dimensional metal: the density of states, the
distribution of Fermi velocities, and the collective electronic response.
\end{abstract}

\begin{multicols}{2}

\begin{figure}

  \centerline{
  \rotatebox{0}{\resizebox{!}{2.2in}{
                \includegraphics{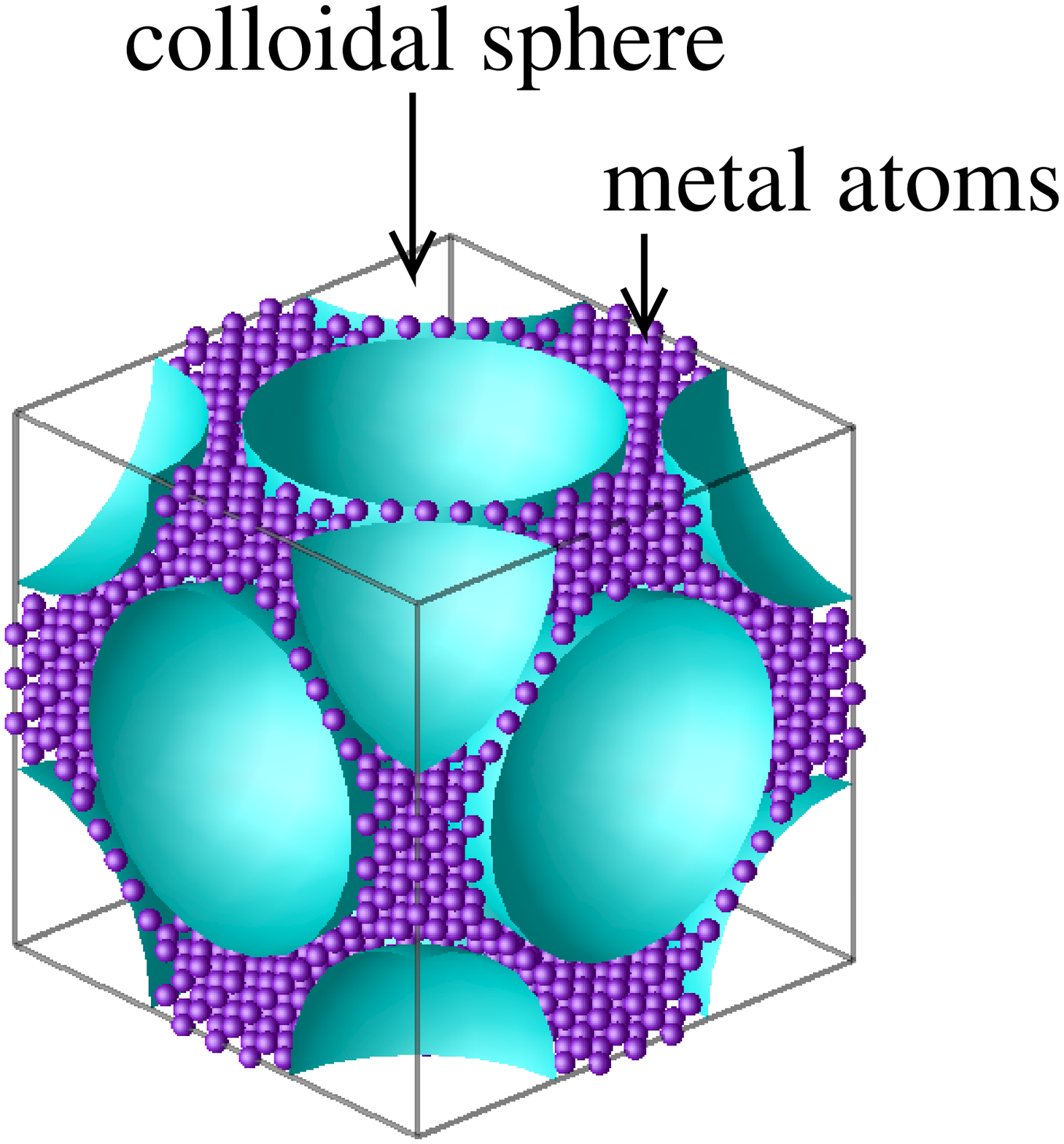}}}}
  \caption{\label{fig1} 
        A face-centered-cubic lattice of insulating colloidal spheres which
        has been infiltrated with (much smaller) metal atoms.
        The colloidal spheres are shown as nearly touching with
        a volume ratio of $V_{Al}/V_{total}$=27\%; they contain a $12\times 12\times 12$ 
        three dimensional hierarchical superlattice, or meta-lattice, of metal atoms. 
        The cross-section of a neck (not easily visible above)
        contains about 10 metal atoms.
        }

  \centerline{
  \rotatebox{-90}{\resizebox{!}{2.5in}{
                \includegraphics{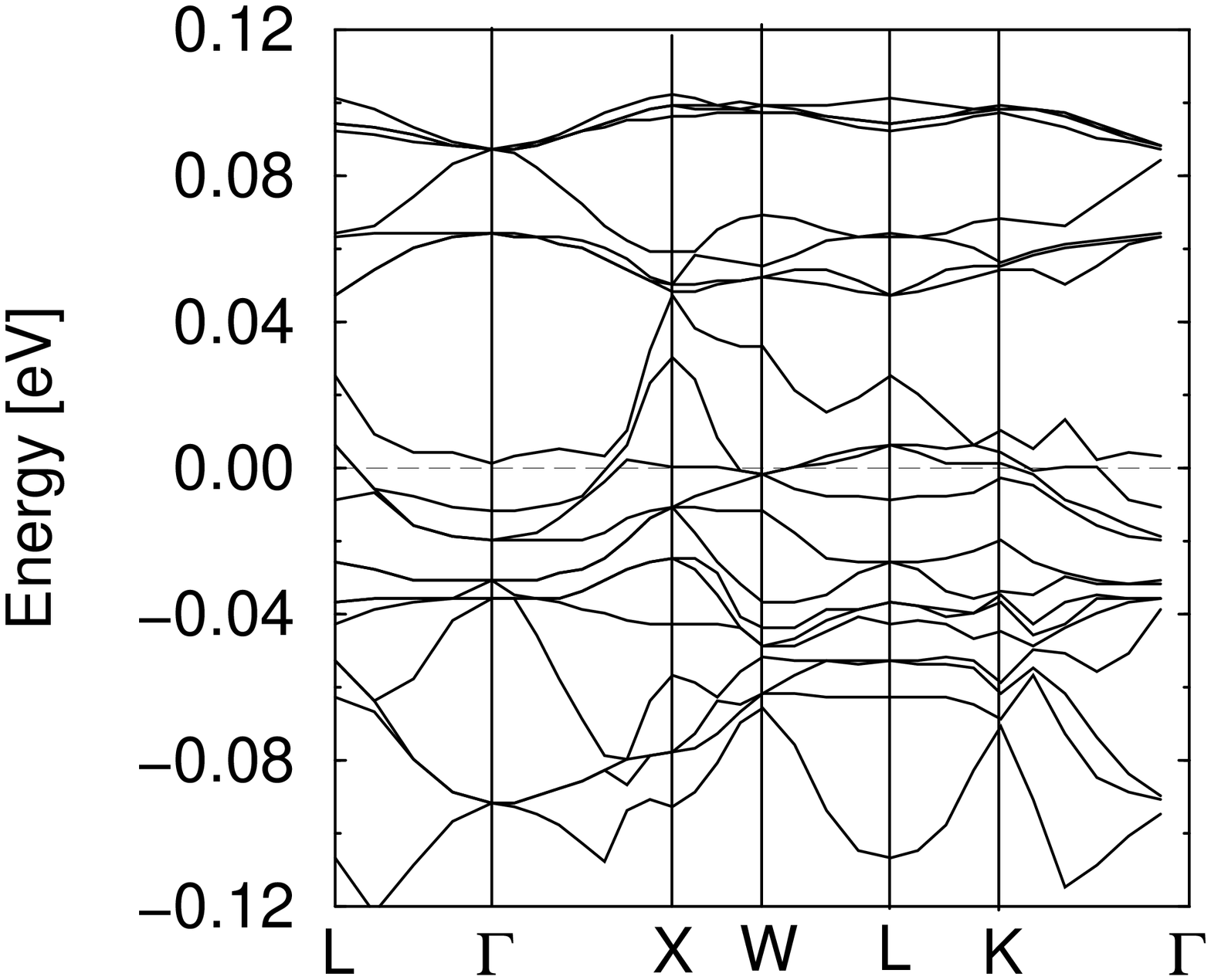}}}}
  \caption{\label{fig2} 
        Band structure along symmetry directions
        for a $9\times 9\times 9$ meta-lattice with touching
        colloidal spheres. Localized states over energies {\it e.g.}
        $0.0\leftrightarrow -0.05$ and $0.08\leftrightarrow 0.1$
        eV come from interior sites in Al clusters. Delocalized states
        interconnecting localized states have significant density on surface 
        and neck sites. The Fermi energy is set
        to 0 eV.}

  \centerline{
  \rotatebox{0}{\resizebox{!}{2.2in}{
                \includegraphics{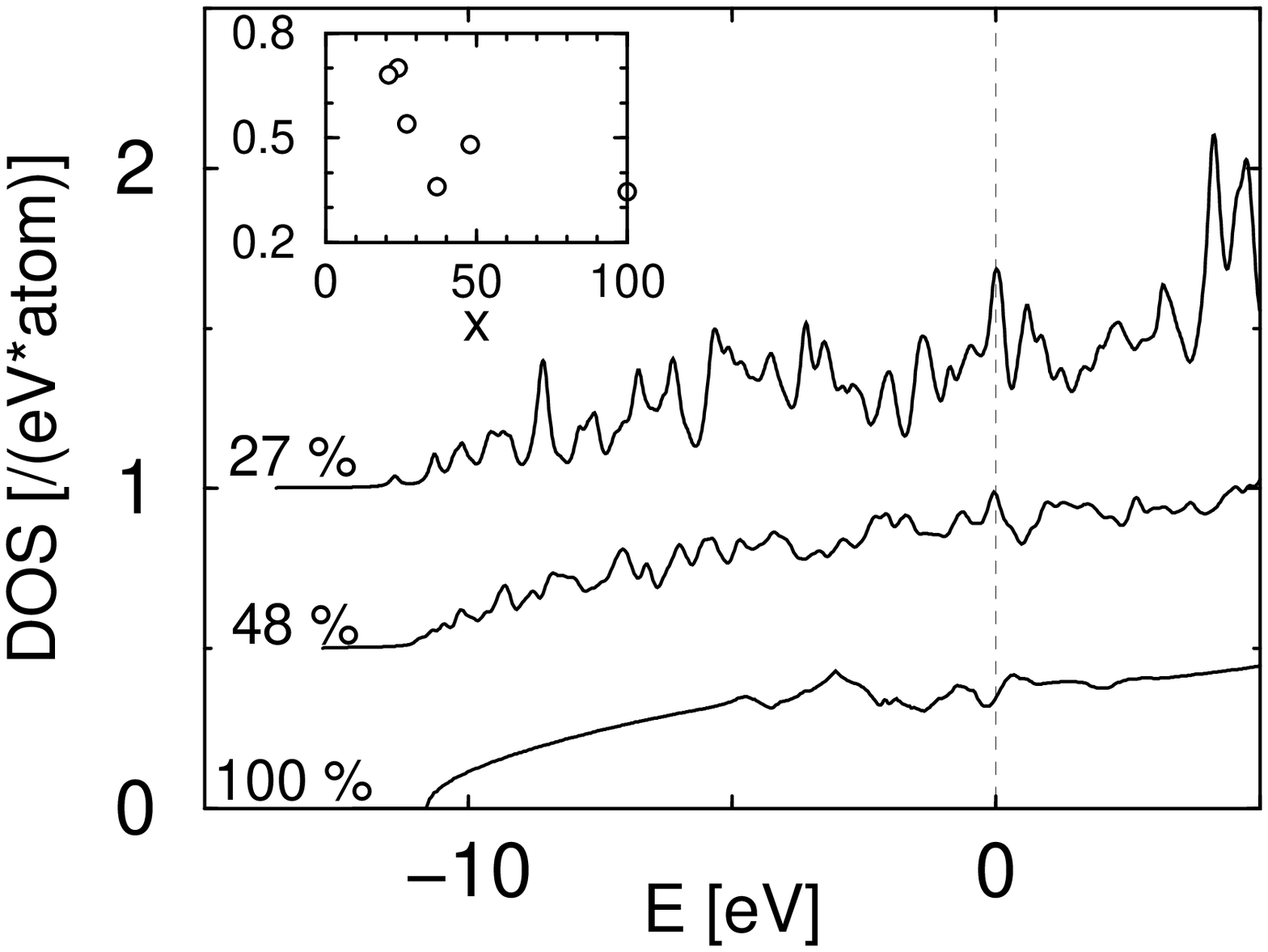}}}}
  \caption{\label{fig3}
        Density of states for varying colloidal radii for a $9\times 9\times 9$ 
        meta-lattice. The main plot shows
        the emerging peaks  with decreasing volume ratio
        $x\ (\equiv V_{Al}/V_{tot})$. With touching colloidal spheres ($x=27$
        \%), the DOS peaks due to nearly localized states.
        The DOS at the Fermi energy ($E_F=0$) tends to
        increase with increasing $x$ (inset).}

  \centerline{
  \rotatebox{0}{\resizebox{3.2in}{!}{
                \includegraphics{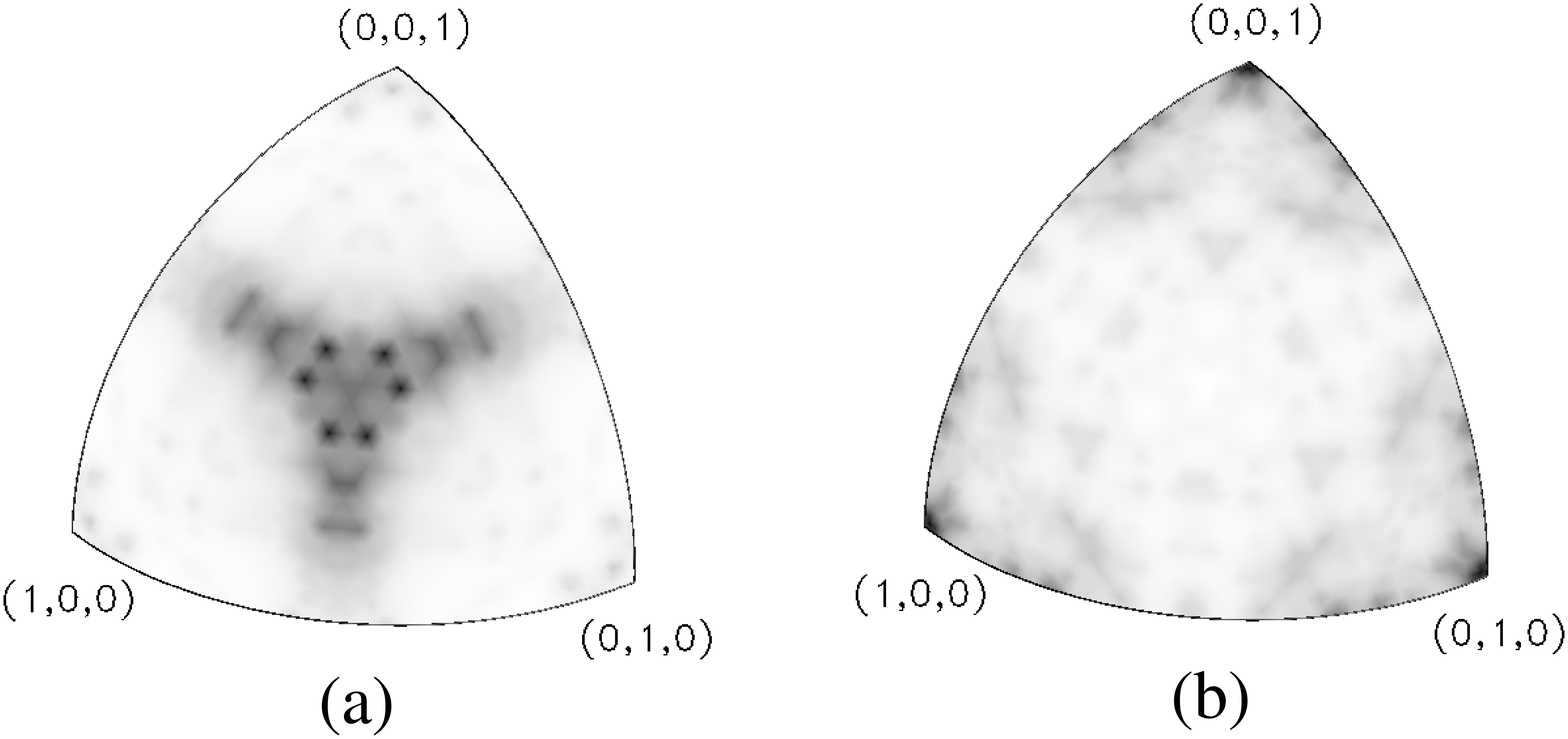}}}}
  \caption{\label{fig4} 
        (a) Bulk aluminum Fermi velocities 
        projected onto a unit sphere with the intensity proportional to
        the magnitude of $v_F$.
        The Fermi velocity is largest along the body diagonal
        direction.
        (b) A similar plot for a $9\times 9\times 9$ meta-lattice with
	touching colloidal spheres: the shadow of (a).
        Dispersive bands contribute to the Fermi velocities along (1,0,0) 
	directions. The Fermi
        velocity along $(0,0,0)$ is strongly suppressed 
        due to intervening colloidal particles.
        }

  \centerline{
  \rotatebox{-90}{\resizebox{!}{2.5in}{
                \includegraphics{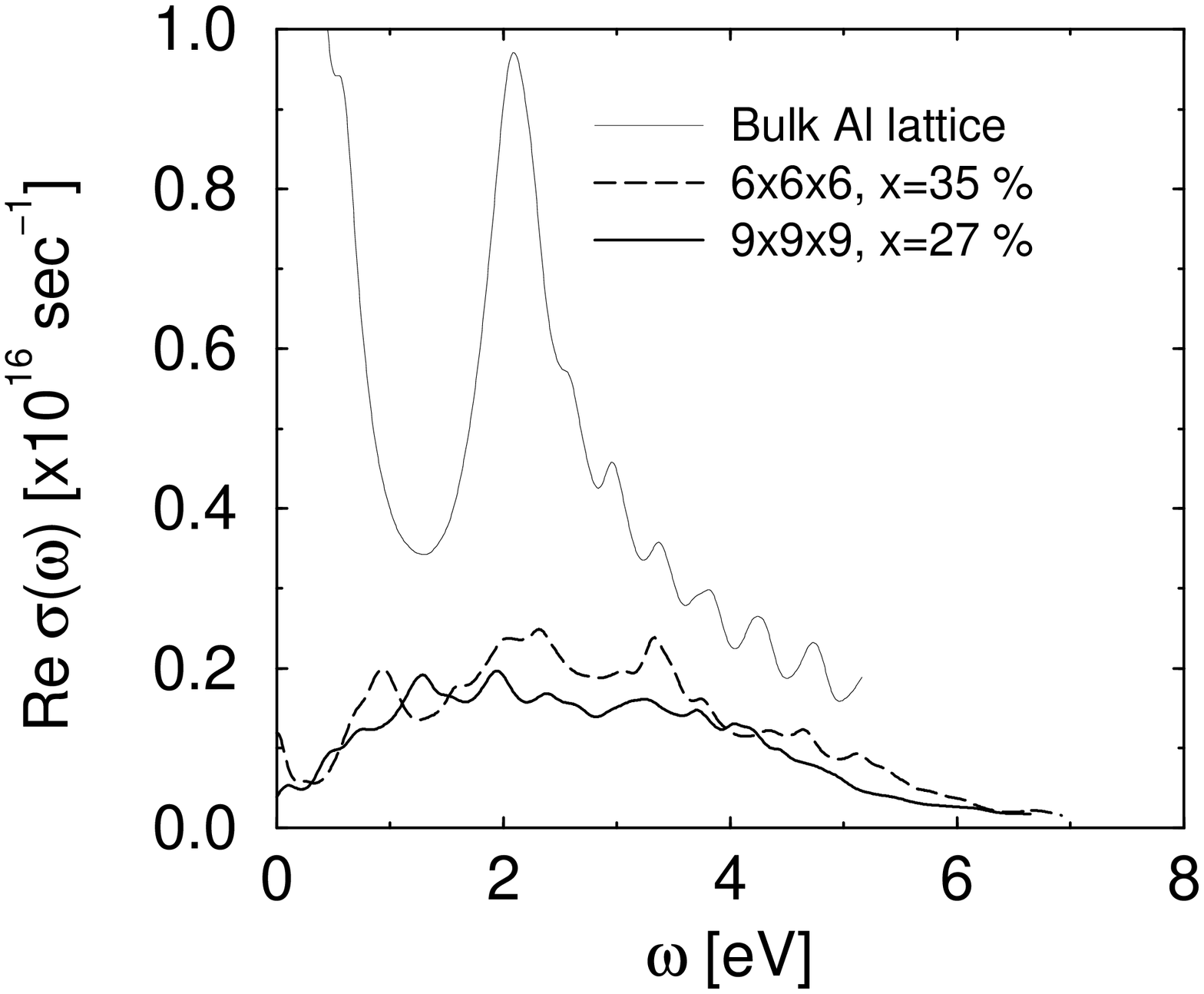}}}}
  \caption{\label{fig5}
        Optical conductivity for a bulk Al lattice and meta-lattice
        with scattering rate $\hbar/\tau=0.1$ eV. In the colloidal
        lattice the intraband contribution, (i.e.\ the Drude part), 
	is strongly reduced
        due to non-dispersive states. The oscillating tail for bulk Al
	is an artifact of finite {\bf k}-mesh.
        }

\end{figure}

Quantum confinement in zero, one and two dimensional semiconductor and
metallic nanostructures has enabled an enormous range of condensed
matter physics and materials science over the past two decades.  The
basic electronic property of a semiconductor, the bandgap, can be
tuned continuously via quantum confinement in nanoparticles. In
contrast, the fundamental defining property of a metallic Fermi
liquid, the Fermi surface, is lost in a quantum dot, since confinement
creates nondispersive electronic states. Although dispersive
low-energy states can be
recovered in low dimensional metals such as nanowires and
two-dimensional electron gases, the degrees of freedom in the confined
and nonconfined directions typically decouple, so that the direct
effects of quantum confinement are restricted to the confined
dimension(s) and often have relatively minor effects on the
single-particle dispersion along the non-confined direction.  Here we
propose a new class of materials that combine quantum confinement with
a continuum of three-dimensional delocalized, dispersive electronic
states through the infiltration of metal atoms into a
three-dimensional colloidal lattice (see Fig. \ref{fig1}).  The
resulting system has a hierarchy of periodicities: the underlying
atomic structure resides within the larger-scale periodic colloid;
hence the term ``meta-lattice.'' This meta-lattice exploits finite
size effects to tune the fundamental low energy ({\it i.e.,} near Fermi 
energy) properties of the metallic state.

Colloidal crystals on the scale of optical wavelengths have been used
as scaffolds for the infiltration of dielectrics into the ordered pore
structure to create photonic materials. Here we consider a new regime
by extending these colloids to the 1 --- 10 nanometer scale and
studying infiltration with metals. Electrochemical deposition,
pressure-induced filling, and iterative chemical reduction of metallic
salts are all being exploited to infiltrate colloidal systems with
metal atoms.  Such systems are just starting to be produced at
nanometer length scales: for example, Pt has recently been infiltrated
into colloidal crystals composed of 30 nm silica spheres with
prospects for extension to smaller sizes\cite{mallouk}.  As yet, no
theoretical treatment exists for the electronic properties of these
ordered nanoscale infiltrant metals.  We demonstrate here that these
three-dimensional metallic meta-lattices induce several new physical
effects such as: (1) Fermi velocity {\it shadowing}, wherein the
ordered interconnecting necks within the structure induce pronounced
modifications into the distribution of Fermi velocities; (2)
Bifurcation of the electronic states into two intimately interspersed
yet distinct ladders of strongly and weakly dispersive states, a
separation which affects the optical response, transport, and the
density of states at the Fermi level; (3) The potential for a rich
interaction of the large-scale (i.e. colloidal) structural order
parameter with other order parameters of the metallic component
(e.g. characteristic superconducting length-scales, etc.).

Each of the effects discussed above applies across a wide class of
metals.  For the purposes of an initial case study, we focus on an
archetypal example: a face-centered-cubic (fcc) lattice of insulating
colloidal spheres, a typical closed packed structure, which is
infiltrated with aluminum (see Fig. \ref{fig1}).  Although aluminum is
not the easiest metal to infiltrate experimentally, its simple
electronic structure and theoretical tractability make it a useful
``canonical metal''\cite{ashcroft}
that can clearly elucidate the generic features of
the new physics.  We assume that the aluminum atoms also crystalize in
the fcc structure inside of the interstitial regions of the lattice,
similar to the structures of Al clusters\cite{rao}.  For computational
convenience, we have chosen the colloidal and Al lattices to be in
alignment; (but note that our main results arise from the large-scale
geometry of three-dimensional confinement, not the detailed atomic-level 
structure
of metal). In this initial investigation we do not explicitly treat
surface reconstructions, which are in any case much less extensive in
metallic systems than in {\it e.g.}  covalently-bonded
semiconductors. We describe disorder through both a phenomenological
broadening and also a random variation in the on-site energies.  
Later we
discuss the differential sensitivity of various phenomena to disorder.

We compute the electronic structure with an empirical $s, p,
d$-orbital tight-binding method whose parameters are obtained by
fitting to the bulk band structures of W\"urde {\it et
al.}\cite{wuerde}. We construct the systems of study by removing, from
the bulk, Al atoms that fall within the colloidal particles.
The resulting aluminum structure then comprises octahedral and
tetrahedral Al islands interconnected by thin metallic necks.  The
volume ratio of aluminum, $x\ (\equiv V_{Al}/V_{tot})$ is roughly
$1-{16\pi\over 3} (R/L)^3$, with $R$ the radius of colloidal spheres
and $L$ the lattice constant of meta-lattice cubic unit cell. In the
physically relevant limit of touching colloidal spheres, $x\approx
0.26$ with $R=L/2\sqrt{2}$. For computational reasons, we restrict the
study to systems up to the $12\times 12\times 12$ meta-lattice,
although many of the effects discussed will persist up to larger
sizes.

Remarkably, instead of narrowing uniformly across all bands, the electronic
structure forms sets of very weakly dispersive states which are
interconnected by more strongly dispersive bands.  Fig. \ref{fig2} shows
the electronic structure around the Fermi energy for an $L= 3.6$ nm
colloidal lattice containing a $9 \times 9 \times 9$ Al meta-lattice at
$x=0.27$. The weakly dispersive states occupying regions around e.g.\ $-0.05
\leftrightarrow 0.00$ and $0.08 \leftrightarrow 0.1$ eV are interconnected
by bands of substantially more dispersive states. This classification
is also seen in real space: the weakly dispersive states predominately
occupy the interiors of the tetrahedral and octahedral islands,
whereas the dispersive states occupy the surfaces and interconnecting
necks.  The weakly dispersive states evolve smoothly into the
localized states of an isolated quantum dot as the lattice
connectivity is reduced through e.g. expanding the colloidal spheres
(theoretically) until they interpenetrate and cut off the necks.
Since the islands contain many more atoms than the necks, most
electrons propagating in an island will reflect from the necks and
thereby acquire only weak dispersion. The weakly dispersive states
(which are likely to localize under disorder) then form a set of
broadened levels with a granularity which is affected by the size of
the islands and the point symmetry of the meta-lattice. In contrast,
the more strongly dispersive states form a new class of confined
electronic wavefunctions which do not have an analogue in quantum
dots.

This granularity in the weakly dispersive manifold produces strong
modulations in the density of states at the Fermi level, even under
the influence of a moderate ($\sim 0.2$ eV) phenomenological
disorder-induced broadening\cite{broadening}.  The main plot in
Fig. \ref{fig3} illustrates this effect by showing the evolution of
the density of states as colloidal spheres of variable size but fixed
spacing are introduced into the meta-lattice.  (Note that the
intermediate values of $x$ are not experimentally accessible, since
the spheres are not touching; instead they provide a well-defined
means to understand the effects of variable quantum confinement at a
fixed colloidal lattice constant).  The calculation covers 89 {\bf
k}-points in the irreducible first Brillouin zone with a Lorentzian
broadening of 0.2 eV (full width at half maximum) to describe disorder
and thermal smearing. The familiar nearly free electron-like density
of states of bulk aluminum (bottom curve) is only slightly changed by
the introduction of periodic voids in the Al lattice at a volume
density of $1-x=0.52$. However, when the colloidal spheres touch at
$x=0.27$, the density of states changes significantly and displays
peaks (arising from the weakly dispersive states) around
$E_F$. Statistically, the Fermi energy tends to fall near such a peak
because such states represent the majority of available states
(i.e. the necks compose only a small fraction of the total atoms) and
a random filling of the bandstructure will tend to locate the Fermi
level in a peak (we neglect the effects of structural relaxation
here).  The inset shows the variation in the Fermi-level density of
states as the volume filling fraction is decreased from the bulk
lattice ($x=1.0$) to the case of touching colloidal spheres
($x=0.27$). As expected, the density of states shows a general trend
towards higher values at smaller volume fraction. The effect, even
under significant ($\sim$ 0.2 eV) disorder broadening, can be
dramatic: at $x=0.27$ the density of states at the Fermi level is more
than twice the bulk value. The specific width of disorder broadening
will depend on the detailed characteristics of any given sample;
regardless, it is surprising and encouraging that the granularity can
survive a broadening which is many times the level spacing expected
for quantum dots of similar size.

The bifurcation of the bulk electronic structure into the weakly and
strongly dispersive states of the meta-lattice also has pronounced
consequences for the collective electronic response.  The plasma frequency,
$\omega_p$, can be calculated from the sum rule for the optical
conductivity $\sigma(\omega)$:
\begin{equation}
\int_0^\infty \sigma(\omega)d\omega = \omega_p^2/8,
\end{equation}
where $\omega_p^2=4\pi ne^2/m^*$ with $n$ the charge carrier density
in a naive kinetic model.  For the bulk lattice, we obtain $w_p=16.3$
eV, in good agreement with experiment\cite{kittel}. The infiltrated Al
meta-lattice has a much reduced plasma frequency: 7.2 and 6.6 eV for
$6\times 6\times 6$ and $9\times 9\times 9$ meta-lattices,
respectively. The ratio of effective charge carriers, $n_{\rm
meta-lat}/n_{\rm bulk}=\omega_{p,\rm meta-lat}^2/\omega_{p,\rm
bulk}^2=0.19$ and $0.16$ respectively, cannot be ascribed solely to
the reduced volume fractions of $x=0.35$ and 0.27.  Instead, the large
reduction demonstrates that roughly half of the charge carriers in the
infiltrated metal do not participate significantly in this aspect of
the collective optical response due to a strongly enhanced effective
mass. Since the two classes of electronic states are intimately
interspersed in energy, they are unlikely to decouple into distinct plasmons,
as would be the case for e.g. $\sigma$ and $\pi$ plasmons in carbon.

The strongly anisotropic lattice connectivity within the infiltrated
colloid also causes qualitative changes to the pattern of Fermi velocities
around the Fermi surface of the metal infiltrant. Fig. \ref{fig4}(a) shows
the Fermi velocities for bulk Al projected onto a unit sphere.  We have
weighted each Fermi velocity vector with its magnitude and then smeared
slightly to obtain a continuous distribution.  The Fermi velocity of bulk
Al is largest (dark region) along the body-diagonal direction.  In
contrast, as shown in Fig. \ref{fig4}(b), the Fermi velocities in the
infiltrated $9\times 9\times 9$ Al meta-lattice concentrate along (1,0,0)
directions.  The remarkable inversion, {\it i.e. shadowing}, in the
distribution of Fermi velocities between the bulk and colloidal lattices
arises because the electron transport in the infiltrated metal is greatly
hindered by the large unpenetrable colloidal spheres which suppress the
Fermi velocity strongly along the body-diagonal directions.  The
non-dispersive states are effectively filtered out from electron transport
and the dispersive states dominate the Fermi velocity distribution.  Since
the dispersive states predominately occupy the surface and necks, the
symmetry of the Fermi velocity distribution is primarily determined by the
overall connectivity of the metal surface, although its details are
sensitive to the atomic-scale structure of the surface.  We expect these
qualitative features to hold regardless of lattice alignments or the
lattice structure of metal infiltrant so long as the size of the colloidal
lattice is smaller than relevant length-scales in the metal (e.g. the
mean free path).

Interesting phenomena also appear in the optical conductivity, even
within a simplified isotropic phenomenological model of the
scattering (see Fig. \ref{fig5}). 
For simplicity, we choose a uniform scattering rate
$1/\tau=0.1$ eV; this choice isolates the effects of the modified
Fermi-level states from those due to possible
confinement/surface-induced modulations in scattering
processes\cite{scattering}. (A more complete treatment of scattering
might reveal new structure in the frequency dependent
conductivity as e.g.\ $v_F/\omega$ approaches the colloidal lattice
constant). For the bulk lattice, the intraband Drude peak dominates,
while the first two interband transitions (the first one is buried
under the Drude peak) give additional structure which is also in good
agreement with experiment\cite{ashcroft,ehrenreich}.  
Despite the large number of states, the
weakly dispersive states contribute only weakly to the optical
conductivity due to their small Fermi velocity.
Transitions between the dispersive states dominate the optical 
conductivity. 

These new phenomena in the bulk-confined ordered metallic state have
variable sensitivity to disorder, including atomic-scale
displacements, variations in colloidal particle size, and grain
boundaries within the metal or colloidal template. The detailed
structure of particular bands as presented here is the feature most
sensitive to disorder, as fluctuations on the order of the (folded)
bandwidth are likely inescapable in the real systems. However, the
coarsened separation of electronic states into weakly and strongly
dispersive states persists over the entire multi-eV bulk
(i.e. unfolded) bandwidth and the density of state modulations can
survive a broadening greatly in excess of the band spacing. As this
classification relates to the differential occupation of island and
neck sites, it is more robust than the details of any single specific
band.  Similarly, the shadowing of the Fermi velocity distribution
arises directly from the anisotropic meta-lattice connectivity, which
is also relatively robust under atomic-scale disorder.
Finite grain size will smear the
Fermi surface, but the Fermi velocity symmetry and weakly/strongly
dispersive character of the states exists over the entire Brillouin
zone. Note that, as a function of colloidal lattice period, different
properties approach the bulk limit at different sizes: the long-range
spatial coherence over a single band state is destroyed by very weak
disorder in larger-scale systems, whereas the density of states
fluctuations can survive to significantly larger sizes (or similarly,
larger disorder smearings), and the shadowing in the angular
distribution of the electronic dispersion can also persist to larger
colloidal-lattice periodicities.

Particularly intriguing properties emerge as the lattice constant of
the infiltrant (and the associated structural order parameter)
approaches other characteristic length-scales in the metal. For
example, at larger colloidal sizes (several hundred Angstroms and
higher) where many of the atomic-scale electronic modulations
described above will begin to wash out, the collodial lattice
structural order parameter will begin to interact richly with {\it
e.g.}  characteristic superconducting length-scales such as the
penetration depth, with potentially fascinating consequences for the
magnetic response of meta-lattice type I or type II superconductors.
In addition, across a range of length-scales the electron
screening and transport response will be sensitive to the physical and
chemical configuration of the surface and interstitial spaces. Since
the original colloidal lattice can be etched away, this newly
liberated interstitial space within the metallic meta-lattice is
available for surface functionalization or second infiltration, both
of which could be reversible in certain cases (i.e. for a fluid
infiltrant). For example, a backfilling infiltration with a dielectric
medium into an open metallic meta-lattice would modulate the
electron-electron interaction, particularly across the void regions.
The single-particle bandstructure results given here provide a base of
understanding towards further study of this fascinating class of
materials.

We thank P.\ Eklund and T.\ Mallouk for help discussions. We gratefully
acknowledge the David and Lucile Packard Foundation, the National Science
Foundation through grant DMR-9876232, and the National Partnership for
Advanced Computational Infrastructure and the Pittsburgh Supercomputing
Center for computational support.


\end{multicols}


\begin{references}
\bibitem{scattering} 
  The scattering times in the bulk and meta-lattice should differ.
  The interband transitions, nevertheless,
  are not as sensitive to the scattering time as are the intraband contribution.
\end{references}
\end{document}